\documentclass[letterpaper, 10 pt, conference]{ieeeconf}  

\usepackage[caption=false,font=footnotesize]{subfig}
\usepackage{amssymb,amsmath,color,theorem}
\usepackage[english]{babel}
\usepackage{graphicx,epstopdf}
\usepackage{placeins}
\usepackage{subcaption}
\usepackage{booktabs}
\usepackage{hyperref}

\newtheorem{theorem}{Theorem}

\usepackage{comment}
\usepackage{mathtools}

\IEEEoverridecommandlockouts                              

\usepackage{amsmath} 
\usepackage{amssymb}  

\title{\LARGE \bf
Physics-Informed Neural Networks for Nonlinear Output Regulation
}

\author{Sebastiano Mengozzi\textsuperscript{1,\dag}, Giovanni B. Esposito\textsuperscript{1,\dag},  Michelangelo Bin\textsuperscript{1}, \\ Andrea Acquaviva\textsuperscript{1}, Andrea Bartolini\textsuperscript{1}, Lorenzo Marconi\textsuperscript{1}
\thanks{The activity of Giovanni B. Esposito, Sebastiano Mengozzi, Andrea Acquaviva, and Andrea Bartolini is supported by the Horizon Europe DECICE Project (g.a. 101092582) and Cineca.
The activity of Lorenzo Marconi and Michelangelo Bin is supported by the national projects PRIN2020 DOCEAT, PRIN2022 ASTRA and the regional project AGRICOBOT.}
\thanks{\textsuperscript{1}Department of Electrical, Electronic and Information Engineering - DEI, University of Bologna, Viale del Risorgimento 2, Italy
        {\tt\small sebastiano.mengozzi@unibo.it, g.esposito@unibo.it}}%
\thanks{\textsuperscript{\dag} These authors contributed equally to this work}
}

\begin{document}

\maketitle
\thispagestyle{empty}
\pagestyle{empty}

\begin{abstract}
This work addresses the full-information output regulation problem for nonlinear systems, assuming the states of both the plant and the exosystem are known. In this setting, perfect tracking or rejection is achieved by constructing a zero-regulation-error manifold \(\pi(w)\) and a feedforward input \(c(w)\) that render such manifold invariant. The pair \((\pi(w),c(w))\) is characterized by the regulator equations, i.e., a system of PDEs with an algebraic constraint.
 We focus on accurately solving the regulator equations by introducing a physics-informed neural network (PINN) approach that directly approximates \(\pi(w)\) and \(c(w)\) by minimizing the residuals under boundary and feasibility conditions, without requiring precomputed trajectories or labeled data. The learned operator maps exosystem states to steady state plant states and inputs, enables real-time computation and, critically, generalizes across families of the exosystem with varying initial conditions and parameters.
The framework is validated on a regulation task that synchronizes a helicopter’s vertical dynamics with a harmonically oscillating platform. The resulting PINN-based controller reconstructs the zero-error manifold with high fidelity and sustains regulation performance under exosystem variations, highlighting the potential of learning-enabled solvers for nonlinear output regulation. The proposed approach is broadly applicable to nonlinear systems that admit a solution to the output regulation problem.
\end{abstract}


\section{INTRODUCTION}

\label{sec:i}
Designing a feedback control law for a plant such that its output asymptotically tracks prescribed trajectories or rejects disturbances is a central problem in control theory.
For example, tracking is crucial when stabilizing a drone in gusty winds or landing a helicopter on a floating platform affected by wave-induced vertical motions \cite{bisheban2020geometric, wang2021backstepping, isidori2001computation, koo1998output}.
This problem is called \emph{output regulation} problem and, for the class of linear systems, it was studied in \cite{davison2003multivariable, francis1976internal} and is related to the solvability of two linear matrix equations. For the class of nonlinear systems, such problems are extremely difficult to solve due to the fact that designing this controller is tied to the solvability of a set of partial differential equations and nonlinear algebraic equations, the so-called Regulator Equations (REs) or Isidori and Byrnes equations \cite{isidori_output_1990}. Solving this problem would benefit a broad class of nonlinear systems that cannot be handled by popular inversion approaches \cite{wang2001neural}. 
Hence, some methods have been introduced to find approximate solutions of these equations, e.g. Galerkin expansion \cite{khailaie2011analytic} and Taylor series \cite{huang2002approximation, huang1994robust}. However, Galerkin expansion methods scale poorly with increasing nonlinearity, and there is no clear rule for selecting the series order to meet accuracy targets.
On the other hand, Taylor-based approximations are valid only in a small neighborhood of the origin and require computing high-order multivariate expansions of several nonlinear terms.
\\
To address these challenges, numerical approximate solutions based on neural networks have been proposed in \cite{wang2001neural, wang2000approximate}. The strength of the neural network approach lies in the fact they can be executed really efficiently with modern hardware accelerators, even in resource-constrained systems, enabling the possibility of deployment in real-world cyber-physical systems, such as drones, helicopters, and robotics systems in general. Moreover, they can approximate the solution of the REs up to an arbitrarily small error in any given compact subset.
However, the approaches proposed in \cite{wang2001neural, wang2000approximate} were developed prior to the widespread adoption of modern AI paradigms and scalable accelerator hardware; consequently, training emphasized data fitting with shallow models and simple optimizers. Modern advances in deep learning and optimization now enable more efficient and accurate techniques such as auto-differentiation to impose physics-informed losses and improve the accuracy of the solution. 

In this work, we employ Physics-Informed Neural Networks (PINNs) \cite{raissi2019physics} to solve the nonlinear REs. 
PINNs and operator learning \cite{kovachki2023neural, li2024physics} have proved effective at solving PDEs across domains such as physics, biology, and fluid dynamics, offering a mesh-free alternative to traditional methods.
Rather than relying on a dataset, which can be costly and impractical, we train PINNs directly from the nonlinear REs, enforcing the PDE residuals and boundary conditions at sampled exosystem states. During the training phase we varied the exosystem initial conditions and parameters, so that the network learns an operator instead of a case-specific solver, yielding steady state solutions for a family of dynamical systems.
In this way, the approach exploits rich a priori knowledge through automatic differentiation, moving beyond black-box learning while still leveraging deep neural networks as universal approximators.
The proposed methodology applies broadly to nonlinear systems that admit an output-regulation solution.
\\
The resulting PINN is a fully connected neural network with four hidden layers and approximately $8\times 10^4$ trainable parameters, assumed to be sufficiently large, and designed to keep inference cost low and enable real-time deployment. We validate the framework on a nonlinear helicopter vertical-tracking benchmark in the full-information setting, showing that the learned operator reconstructs the zero-error manifold and maintains regulation performance across a wide range of exosystem initial conditions and frequencies, including configurations not seen during training.
The proposed framework reconciles the theoretical structure of output regulation with learning-based methods, providing a principled path toward data-efficient, learning-based controllers for nonlinear output regulation problems.

\section{PRELIMINARIES}
\paragraph*{Notation}
Scalars, mappings, vectors, and vector
fields are denoted by lowercase italic letters while constant parameters by uppercase italic letters; matrices are denoted by uppercase bold letters; calligraphic letters denote sets. The Lie derivative along the exosystem vector field $s(\cdot)$ is denoted by $\mathcal{L}_s$, and the loss function by $\mathcal{J}$. Mechanical vectors are denoted by bold lowercase letters.  Explicit time dependence is omitted for simplicity, whereas dependence on the exosystem state $w$ is indicated explicitly. Hatted symbols denote approximations or learned quantities. 

\subsection{The Output Regulation Problem for Nonlinear Systems}

\label{sec:ii}
We consider the smooth, time-invariant plant--exosystem interconnection
\begin{equation}
\label{eq:plant}
\begin{aligned}
    \dot{x} &= f(x,u,w), \\
    e &= h(x,w)
\end{aligned}
\end{equation}
where $x \in \mathbb{R}^{n}$ is the plant state, $u \in \mathbb{R}^{m}$ the control input, and $e \in \mathbb{R}^{p}$ the regulation error. The exogenous signal $w \in \mathbb{R}^{s}$ evolves according to the autonomous exosystem
\begin{equation}
\label{eq:w}
\dot{w} = s(w).
\end{equation}
We adopt the \emph{full-information} setting \cite{isidori_output_1990}, meaning that both $x$ and $w$ are available. The goal is to achieve $e=h(x,w) \to 0$ while the exosystem evolves according to $s$. In steady state, perfect regulation is described by two mappings $x=\pi(w)$ and $u=c(w)$ that parameterize a \emph{zero-error manifold} and the corresponding feedforward input that keeps the motion on that manifold. Substituting these maps into~\eqref{eq:plant} and imposing $e\equiv 0$ yields the \emph{regulator equations}, a system of PDEs with an algebraic constraint:
\begin{equation}
\label{eq:regulator}
\begin{aligned}
   \mathcal{L}_{s}\pi(w) &= f\big(\pi(w),\,c(w),\,w\big), \\
    0 &= h\big(\pi(w),\,w\big),
\end{aligned}
\end{equation}
where $\mathcal{L}_{s}\pi(w):=\frac{\partial \pi}{\partial w}(w)\,s(w)$. Intuitively, $\pi(w)$ collects steady plant states consistent with the exogenous signal, while $c(w)$ renders the graph of $\pi$ invariant. For well-posedness, one typically supplements~\eqref{eq:regulator} with anchoring and feasibility conditions (e.g., $\pi(0)=0$, $h(\pi(0),0)=0$) and seeks solutions in a neighborhood of the origin. A key implication of \eqref{eq:regulator} is that the \emph{structure} of the exosystem directly enters the differential operator $\mathcal{L}_{s}\pi(w)$; consequently, changes in the exosystem --- for instance, a harmonic oscillator with \emph{varying frequency} $\Omega$ --- induce a \emph{family} of PDEs parameterized by $\Omega$. In practice, one may either recompute $(\pi,c)$ for each member of such a family or approximate a mapping that generalizes across these variations. Once a candidate pair $(\pi(w),c(w))$ is available, a local stabilizer can be incorporated to make the zero-error manifold attractive,
\begin{equation}
\label{eq:u}
\begin{aligned}
    u &= \mathbf K\big(x-\pi(w)\big) + c(w),
\end{aligned}
\end{equation}
under standard stabilizability assumptions. Here the feedback serves to regularize transients around the invariant manifold, while the steady state behavior is dictated by~\eqref{eq:regulator}. The formulation will be illustrated on a benchmark involving the synchronization of a helicopter's vertical dynamics with a harmonically oscillating platform, which we use to assess generalization with respect to exosystem variations~\cite{isidori2001computation}.

\section{PROPOSED SOLUTION}
\label{sec:iii}
This work develops a methodology to approximate the solutions of the regulator equations \eqref{eq:regulator} building on the PINNs framework. The key idea is to decouple the solution from time: rather than learning the solutions in time, as usually done with PINNs, we directly learn the steady state mappings as functions of the exosystem state $w$. In this formulation, $w$ is a point in a set $\mathcal{W}$ that collects all admissible exosystem states. In other words, $\mathcal{W}$ represents the set of possible inputs the exosystem can generate.
This perspective removes the need to construct a dataset, which is typically costly and cumbersome.
To achieve this, we introduce a parametrized operator $n_{\xi}$, with trainable parameters $\xi$, to approximate the two mappings $\pi(w)$ and $c(w)$ defined in~\eqref{eq:regulator}. The operator is implemented as a feed-forward neural network that takes as input a sampled exosystem state $w_i$ and returns the corresponding steady state approximations $\hat{\pi}(w_i)$ and $\hat{c}(w_i)$:
\begin{equation}
    \big[\hat\pi(w_i), \hat c(w_i)\big] = n_{\xi}(w_i).
\end{equation}
Since the exact solution of~\eqref{eq:regulator} is not available, we assess the quality of these approximations through a PINN-based residual, defined as
\begin{equation}
\begin{aligned}
    \label{eq:residual}
    r_{\mathrm{RE}}(w_i) &= \left\|\mathcal{L}_s\hat\pi(w_i) - f\big(\hat\pi(w_i), \hat c(w_i), w_i\big)\right\|_2 \\
    &\quad + \left\|h\big(\hat\pi(w_i), w_i\big)\right\|_2,
\end{aligned}
\end{equation}
where $r_{\mathrm{RE}}$ is the residual of the regulator equations, i.e., the norm of the mismatch between the left-hand and right-hand sides of the PDE and algebraic constraint evaluated at $(\hat\pi(w^i), \hat c(w^i))$.
The neural operator $n_{\xi}$ is trained by minimizing the loss function
\begin{equation}
    \label{eq:loss}
    \mathcal{J} = \mathcal{J}_{\mathrm{PDE}} + \lambda \cdot \mathcal{J}_{\mathrm{BC}},
\end{equation}
where 
\begin{equation}
    \label{eq:loss_pde}
    \mathcal{J}_{\mathrm{PDE}} = \underset{w\in \mathcal{W}}{\mathbb{E}}[r_{\mathrm{RE}}(w)]
\end{equation}
is the expectation of $r_{\mathrm{RE}}(w)$ over the domain $\mathcal{W}$, and $\mathcal{J}_{\mathrm{BC}}$ is a suitably defined penalty term to enforce the system’s boundary conditions, and $\lambda$ is a parameter controlling the ratio of importance between the two terms.
By construction, this loss is fully unsupervised and does not require any precomputed numerical solution or labeled dataset.
The model is designed to solve a family of PDEs rather than a single instance: by sampling from a family of dynamical systems, the network is trained not to approximate one particular solution, but to act as a solution operator that generalizes to unseen parameter configurations.
The optimization is performed over the network’s weights and biases using automatic differentiation. A distinctive feature of the PINN-based formulation is that automatic differentiation is also applied with respect to the network input to compute the Lie derivatives \(\mathcal{L}_s \pi(w^i)\), which contribute to the residual \(r_{\mathrm{RE}}(w^i)\) and thus to the gradients used for updating the network parameters.

\section{VERTICAL LANDING OF A HELICOPTER}
\label{sec:iv}

In this section we present the nonlinear benchmark mechanical system; we refer to \cite{isidori2001computation, koo1998output} for more details on the model derivation. 
The control problem was selected because it involves an underactuated, multi-input mechanical system with coupled and nonlinear dynamics conditions for which, to the best of our knowledge, no existing method provides an exact solution.

\subsection{Helicopter Model}
The helicopter is modeled as a rigid body, neglecting both
fuselage and rotor flexibility as well as actuation dynamics. The forces and torques produced by the actuators are therefore treated directly as control inputs. These inputs are \(u_{T_M},u_{T_T}, u_a,\) and \(u_b\) representing the main rotor thrust, the tail rotor thrust and the longitudinal and lateral tilt of the tip path plane of the main rotor with respect to the shaft, respectively. The superscript \(^B\) and \(^I\) indicates that the quantity is expressed with respect to the body frame and the inertial frame respectively. Let \(\textbf{f}^B\ \) and \(\boldsymbol{\tau}^B\) be the force and torque acting on the center of the mass of the rigid body, \( \mathbf p^I = [p_x^I, p_y^I, p_z^I]^T\) and \( \mathbf v^I\)  the position vector and the velocity vector of the body frame,  and \(\boldsymbol\omega^B\) its angular velocity.
The system dynamics follow the Newton-Euler equations

\begin{equation}
\label{eq:model}
\begin{aligned}
m \dot{\mathbf v}^I &= \mathbf R(\mathbf q)\mathbf f^B, \qquad\\
\dot{\boldsymbol\omega}^B &= \mathbf J^{-1}\!\left(\boldsymbol\tau^B - \boldsymbol\omega^B \times \mathbf J \boldsymbol\omega^B\right),
\end{aligned}
\end{equation}
where $ m $ is the mass, $ \textbf{J} $ is the inertia matrix, and $ \textbf{R}(\mathbf q) $ is the rotation matrix associated with the roll, pitch, and yaw angles $ \mathbf q = [\psi, \theta, \phi]^T $, which satisfy
\begin{equation}
\begin{aligned}
\label{eq:rpy}
    \dot{\mathbf q} &= \textbf{D}(\mathbf q)\, \boldsymbol\omega^B,
\end{aligned}
\end{equation}
with $\textbf{D}(\mathbf q)$ defined as
\begin{equation}
\begin{aligned}
\label{eq:rot}
    \textbf{D}(\mathbf q) &= \begin{bmatrix} 
    1 & \sin\phi \tan\theta & \cos\phi \tan\theta \\
    0 & \cos\phi & - \sin\phi \\
    0 & \dfrac{sin\phi}{\cos\theta} & \dfrac{cos\phi}{\cos\theta}
    \end{bmatrix} .
\end{aligned}
\end{equation}
Under these assumptions, the total force in the body frame is the sum of the main and tail rotor, respectively $\textbf{f}_{M} = [\text{f}_{M,x}, \text{f}_{M,y}, \text{f}_{M,z}]^T$ and $\textbf{f}_{T} = [\text{f}_{T,x}, \text{f}_{T,y}, \text{f}_{T,z}]^T$, and gravity:

\begin{equation}
    \textbf{f}^B = \begin{bmatrix} - u_{T_M}\sin u_a \\  u_{T_M} \sin u_b -u_{T_T} \\ - u_{T_M}\cos u_a\cos u_b
    \end{bmatrix} + \textbf{R}^T \begin{bmatrix}
        0 \\ 0 \\ mg
    \end{bmatrix}.
\end{equation}
The total torque is given by the sum of rotor torques and the moments induced by the rotor forces $\boldsymbol\tau_\text{f}$:

\begin{equation}
        \boldsymbol{\tau}^B = \begin{bmatrix} c_{b}^{M}u_b - Q_M \sin u_a\\  c_{a}^{M}u_a + Q_M \sin u_b -Q_T \\ -Q_M \cos u_a \cos u_b
    \end{bmatrix} +\begin{bmatrix}
        \tau_{\text{f}, x} \\ \tau_{\text{f}, y} \\ \tau_{\text{f}, z}
    \end{bmatrix},
\end{equation}
with $ Q_M = c_{M}^{Q}(u_{T_{M}})^{\frac{3}{2}} - D_{M}^{Q}$, $
        Q_T = c_{T}^{Q}(u_{T_{T}})^{\frac{3}{2}} - D_{T}^{Q}.$
The force-induced moments \( \boldsymbol\tau_{\text{f}}=[\tau_{\text{f}, x} , \tau_{\text{f}, y} , \tau_{\text{f}, z}]^T\) follow the helicopter geometry:
\begin{equation}
    \begin{aligned}
        \tau_{\text{f},x} &=u_{T_M} \sin u_b h_M -u_{T_M}\cos u_a \cos u_b  y_M -u_{T_T} h_T, \\ 
        \tau_{\text{f},y} &=  u_{T_M}\sin u_a h_M -u_{T_M}\cos u_a \cos u_b l_M, \\ 
        \tau_{\text{f},z} &= -u_{T_M} \sin u_b l_M +u_{T_T} l_T.
    \end{aligned}
\end{equation}
The parameters $c_{b}^{M}$, $ c_{a}^{M}, c_{M}^{Q}$, $c_{T}^{Q}$, $D_{M}^{Q}$, $D_{T}^{Q}$, $h_M$, $h_T$, $l_M$, $l_T$, and $y_M$ are the physical parameters of the helicopter.
The values used in the simulation are reported in the Appendix.

\subsection{Exosystem Model}
In this framework, the exosystem in~\eqref{eq:w} is assumed to be known and thus exploitable in the computation of the zero-error manifold mappings. The specific trajectory \(w(t)\), however, is not fixed: the initial condition \(w(0)\) and the exosystem parameters are unknown but constrained to lie in prescribed ranges. This assumption represents a compromise between the ideal yet unrealistic case in which \(w(t)\) is perfectly known, and the opposite conservative case in which \(w(t)\) is treated as a completely arbitrary signal.
A representative example is when \(w(t)\) belongs to the class of periodic functions with unknown frequency and amplitude. The extension of the analysis to more general periodic signals does not pose any conceptual difficulty.
\\
Given these conditions, we can define the exosystem model as
\begin{equation}
    \begin{aligned}
        \dot{w} = s(w)= \mathbf{S}w =\begin{bmatrix}
            0 & \Omega \\
            -\Omega & 0
        \end{bmatrix} \begin{bmatrix}
            w_1 \\ w_2
        \end{bmatrix}
    \end{aligned}.
\end{equation}
This system generates the time-dependent vertical sinusoidal reference to be tracked, with frequency $\Omega$.

\subsection{Regulation Problem}
The regulation goal is to track the error
\begin{equation}
\label{eq:error}
   \mathbf e = \begin{bmatrix} p_x, & p_y, & p_z - w_1 \end{bmatrix}^T = \mathbf 0
\end{equation}
which corresponds to keeping the helicopter motion on a steady state submanifold parametrized by the mappings
\begin{equation}
    \mathbf p^{I} = \pi_p(w), \; \mathbf v^I = \pi_v(w), \; \mathbf q = \pi_q(w), \; \dot{\mathbf q} = \pi_{\dot{q}}(w),
\end{equation}
and rendered invariant through the steady state control laws
\begin{equation}
    u_{T_M} = c_{T_M}(w), \; u_{T_T} = c_{T_T}(w),\; u_a = c_{a}(w), \; u_b = c_{b}(w).
\end{equation}
Following the derivation in~\cite{isidori2001computation}, we set the steady state roll dynamics to 
\(\pi_\psi(w)\equiv 0\), reducing the problem to determining 
\(\pi_\theta(w)\), \(\pi_\psi(w)\), and the control mapping \(c_b(w)\). 
By expressing the torques \(\boldsymbol\tau^b\) in terms of the control inputs, one obtains the steady state relationships
\begin{equation}
\label{eq:c}
    \begin{aligned}
\tan c_{a}(w) \quad &= \quad-\,\tan \pi_{\theta}(w)\,\frac{\cos c_{b}(w)}{\cos \pi_{\phi}(w)},\\
c_{T_M}(w) \quad &= \quad\frac{\cos \pi_{\phi}(w)\,\cos \pi_{\theta}(w)}{\cos c_{a}(w)\,\cos c_{b}(w)}\,k(w),\\
c_{T_T}(w)  \quad&= \quad c_{T_M}(w)\,\sin c_{b}(w)+\\
&\qquad+ \sin \pi_{\phi}(w)\,\cos \pi_{\theta}(w)\,k(w),
    \end{aligned}
\end{equation}
where \(k(w) \coloneqq m(\Omega^2w_1 +g)\).
Using the kinematic relations~\eqref{eq:rpy}, \eqref{eq:rot}, and \eqref{eq:model} one obtains (we omit dependence on $w$ for compactness): 
\begin{equation}
\label{eq:pidot}
    \begin{aligned}
        \begin{bmatrix}
            \ddot{\pi}_{\phi} \\
            \ddot{\pi}_{\theta} \\
            \dot{\pi}_{\theta}\dot{\pi}_{\phi} 
        \end{bmatrix} &= 
        \begin{bmatrix}
            1 & 0 & 0 \\
            0 & \cos \pi_\phi & \sin\pi_\phi \\
            0 & -\sin \pi_\phi & \cos \pi_\phi
        \end{bmatrix} \cdot \textbf{J}^{-1} \cdot \\
        &\:\:\:\cdot
        \begin{bmatrix}
            \tau^{B}_x - (J_z - J_y)\cos \pi_\phi \sin \pi_\phi \dot{\pi}^{2}_\theta\\
            \tau^{B}_y - (J_x - J_z)\sin \pi_\phi \dot{\pi}_\phi \dot{\pi}_{\theta} \\
            \tau^{B}_z - (J_y - J_x)\cos \pi_\phi \dot{\pi}_\phi \dot{\pi}_{\theta}
        \end{bmatrix}.
    \end{aligned}
\end{equation}
Given the definition of the torque vector $\boldsymbol{\tau}^B = [\tau^B_x,\tau^B_y,\tau^B_z]^T $ and \eqref{eq:c}, equation \eqref{eq:pidot} is a function of \(\pi_\phi(w)\), \(\pi_\theta(w)\), \(c_b(w)\), and \(w\) and can be rewritten compactly as the set of regulator equations
 \begin{equation}
 \label{eq:reheli}
     \begin{aligned}
         \mathcal{L}^2_s\pi_\phi \quad = \quad \mathcal{F}_1(\pi_\phi, \pi_\theta, \mathcal{L}_s\pi_\phi, \mathcal{L}_s\pi_\theta, c_b, w) \\
         \mathcal{L}^2_s\pi_\theta \quad = \quad \mathcal{F}_2(\pi_\phi, \pi_\theta, \mathcal{L}_s\pi_\phi, \mathcal{L}_s\pi_\theta, c_b, w) \\
        0 \quad = \quad \mathcal{F}_3(\pi_\phi, \pi_\theta, \mathcal{L}_s\pi_\phi, \mathcal{L}_s\pi_\theta, c_b, w) \\
     \end{aligned}
 \end{equation}
where \(\mathcal{F}_1\), \(\mathcal{F}_2\), and \(\mathcal{F}_3\) are suitably defined functions. Their solutions \(\pi_\phi(w)\), \(\pi_\theta(w)\), and \(c_b(w)\), together with \(\pi_\psi(w)\equiv 0\) and~\eqref{eq:c}, fully characterize the steady state helicopter motion.
The resulting steady state input vector is $c(w) = [c_{T_M}(w),\, c_{T_T}(w),\, c_a(w),\, c_b(w)]$, 
which ensures that the tracking error~\eqref{eq:error} remains identically zero. The corresponding mapping $
\pi_q(w) = [\pi_\phi(w), \pi_\theta(w), \pi_\psi(w)\equiv 0]$
defines the internal dynamics on the zero-error manifold.
Finally, the admissible operating region is constrained by the boundary condition \cite{koo1998output}
\begin{equation}
\label{eq:boundary}
    |u_b| \le 0.3491,
\end{equation}
with the remaining bounds obtained by substituting this condition into~\eqref{eq:c}.

\subsection{Neural Network Design}
\begin{figure}
    \centering
    \includegraphics[width=1\linewidth]{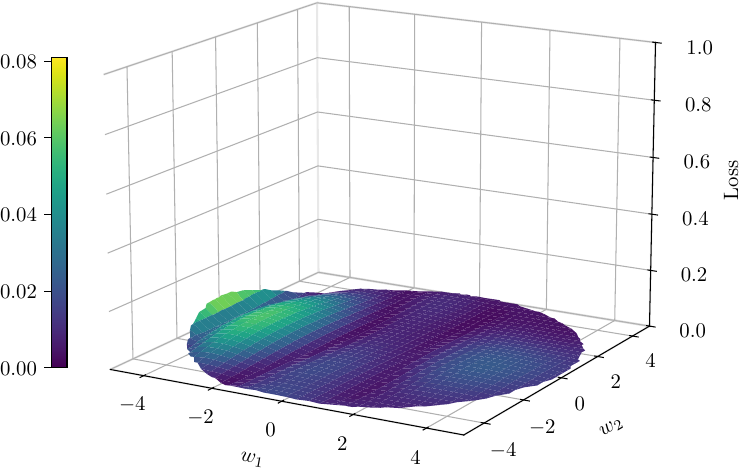}
    \caption{Loss landscape of the trained PINN evaluated over pairs of $(w_1, w_2)$, including both seen and unseen states. The surface is colored according to the loss value.}
    \label{fig:loss3d}
\end{figure}

To approximate the mappings that solve~\eqref{eq:reheli}, we employ a fully connected neural network
\begin{equation}
    n_{\xi}:\mathcal{W} \rightarrow \mathbb{R}^3,
\end{equation}
with parameters $\xi$, which maps exosystem states $w \in \mathcal{W}$ to the three steady state quantities $[\pi_\phi,\, \pi_\theta,\, c_b]$. 
Since $\mathcal{W}$ is continuous, training is performed on a finite subset $\mathcal{W}_K \subset \mathcal{W}$ that is compact and $\varepsilon$-dense.
Each training sample is generated through a polar parameterization $w_i = \big[\, r_i \cos\alpha_i,\; r_i \sin\alpha_i \,\big]$, constructed on a grid of radii and angles. To maintain an approximately uniform spatial density, the number of angular samples increases proportionally with the radius, preventing oversampling near the origin while ensuring adequate coverage at larger radii, we use $
r_i \in \{0, 0.5, 1, \dots, 5.5, 6\}$, with $\alpha_i \in [0, 2\pi]$.
The regulator equations~\eqref{eq:reheli}, together with the boundary conditions~\eqref{eq:boundary}, yield the loss components:
 \begin{equation}
 \label{eq:loss_explicit}
     \begin{aligned}
        \mathcal{L}_{\mathrm{PDE}_1} \  &= \ ||\mathcal{L}^2_s\hat\pi_\phi - \mathcal{F}_1(\hat\pi_\phi, \hat\pi_\theta, \mathcal{L}_s\hat\pi_\phi, \mathcal{L}_s\hat\pi_\theta, \hat c_b, w)||_2 \\
         \mathcal{L}_{\mathrm{PDE}_2} \  &= \ ||\mathcal{L}^2_s\hat\pi_\theta - \mathcal{F}_2(\hat\pi_\phi, \hat\pi_\theta, \mathcal{L}_s\hat\pi_\phi, \mathcal{L}_s\hat\pi_\theta, \hat c_b, w)||_2\\
         \mathcal{L}_{\mathrm{PDE}_3} \  &= \ ||\mathcal{F}_3(\hat\pi_\phi, \hat\pi_\theta, \mathcal{L}_s\hat\pi_\phi, \mathcal{L}_s\hat\pi_\theta, \hat c_b, w)||_2 \\
         \mathcal{L}_{\mathrm{BC}} \ &= \ \max(0, |\hat c_b|-0.3491)
         \\
     \end{aligned}
 \end{equation}
whose sum constitutes the total loss~\eqref{eq:loss}. We empirically set $\lambda=0.1$;  although the boundary constraint was satisfied even with $\lambda = 0$, including this term facilitated early-stage convergence.
Fig.~\ref{fig:loss3d} shows the loss landscape of the trained PINN over pairs $(w_1, w_2)$, including points not seen during training. The loss remains uniformly small across the domain.

\section{EXPERIMENTAL RESULTS}
\label{sec:v}

\begin{figure}
    \centering
    \includegraphics[width=1\linewidth]{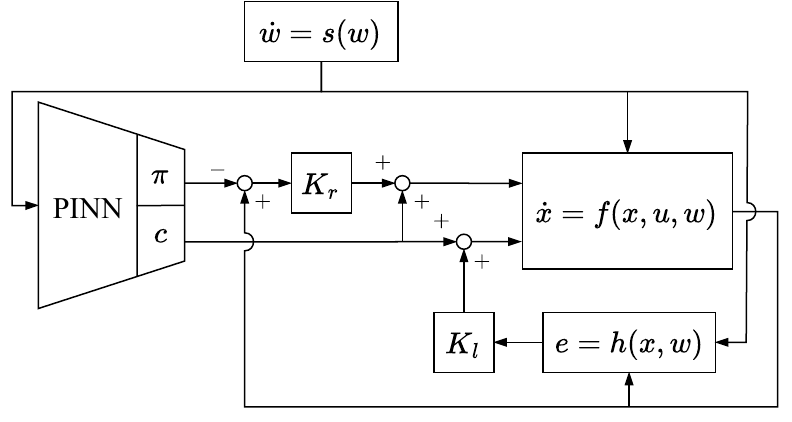}
        \caption{Control structure for the vertical landing problem. The exosystem $\dot{\omega}$ generates the disturbance acting on the plant $\dot{x}$, while the goal is to maintain $e=0$ through the feedforward action and steady plant states produced by the PINN.}
    \label{fig:ctrlschema}
\end{figure}

\begin{figure}[htbp]
    \centering
    \includegraphics[width=1\linewidth]{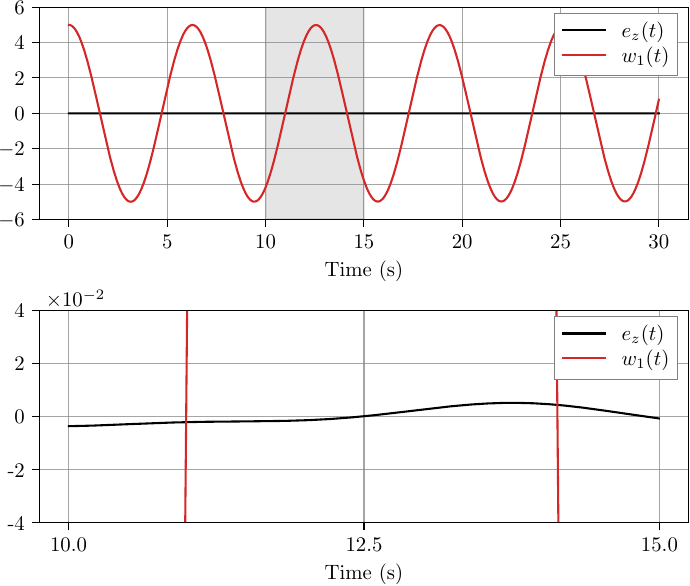}
    \caption{Trajectories of the exosystem reference signal and the vertical tracking error over time in the exosystem configuration $(w_1(0),\Omega) = (5,1)$, which was seen during training. The bottom plot shows a $100\times$ magnification of the error trajectory to highlight the residual ripple.}
    \label{fig:comparison-id}
\end{figure}

\begin{figure}[htbp]
    \centering
    \includegraphics[width=1\linewidth]{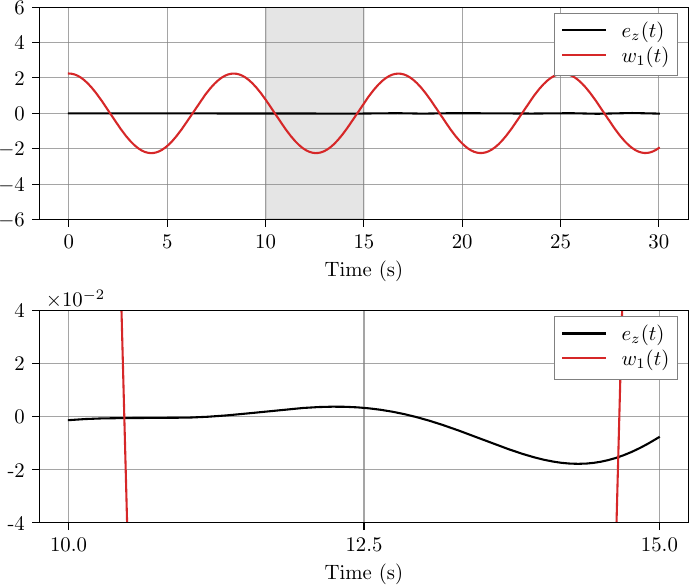}
    \caption{Trajectories of the exosystem reference signal and the vertical tracking error over time in the exosystem configuration $(w_1(0),\Omega) = (2.25,0.75)$, which was \emph{not} seen during training. The bottom plot shows a $100\times$ magnification of the error trajectory to highlight the residual ripple.}
    \label{fig:comparison-ood}
\end{figure}

\begin{figure}[t]
    \centering
        \includegraphics[width=0.8\linewidth]{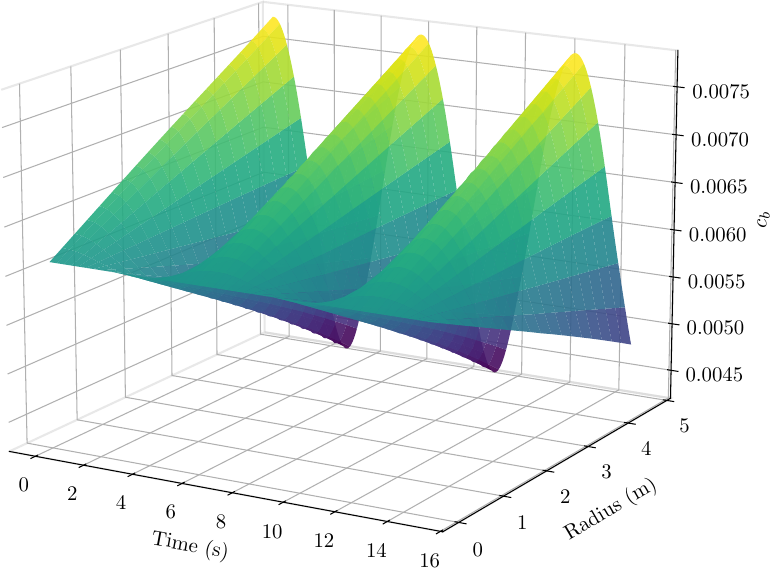}
        \includegraphics[width=0.8\linewidth]{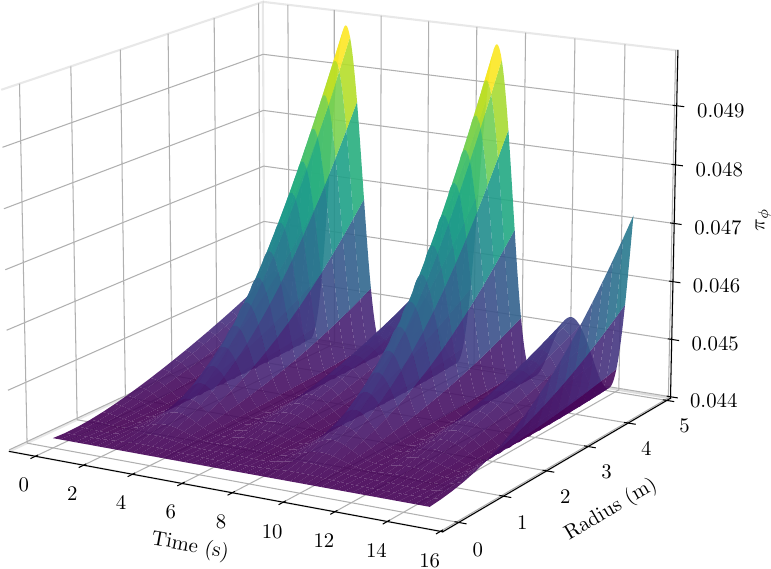}
        \includegraphics[width=0.8\linewidth]{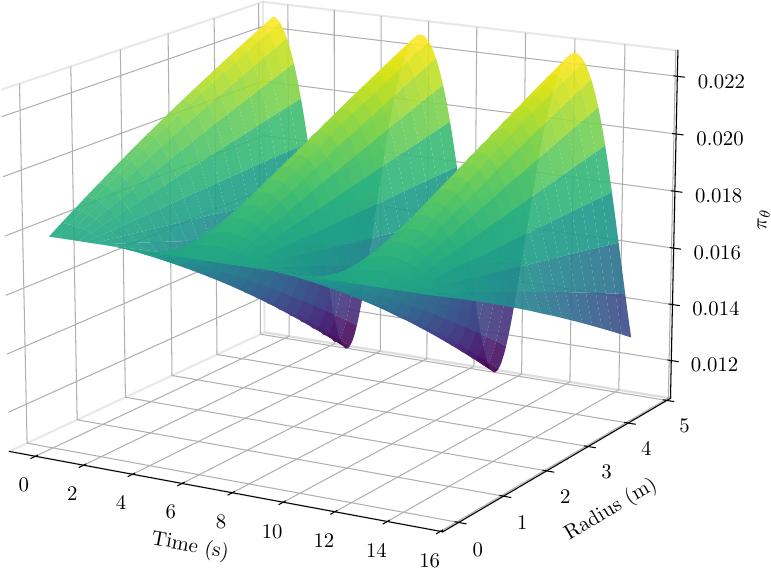}

    \caption{Behavior of the learned mappings $c_b(w)$, $\pi_\phi(w)$, and $\pi_\theta(w)$ as a function of time for different values of the initial condition $w_1(0)$.}
    \label{fig:3maps}
\end{figure}

This section presents the simulation results obtained using the physical parameters reported in the Appendix (see also \cite{koo1998output}).
The proposed PINN-based controller is evaluated on the vertical landing of a helicopter problem described in section \ref{sec:iii}.  The helicopter is initialized on the steady state manifold and simulated for $30\ s$ with a time step of $0.01 \ s$ using the Runge-Kutta integration method.
The control law in~\eqref{eq:u} is implemented as

\begin{equation}
\begin{aligned}
\boldsymbol{\tau}^{B}_{\mathrm{cmd}}
&=
\mathbf{J}\!\Bigl(
\boldsymbol{\tau}^{B}_{\mathrm{ff}}(w)
- K_{r,1}\bigl(\mathbf q-\pi_q(w)\bigr)
- K_{r,2}\,\boldsymbol{\omega}^B
\Bigr) \\
& \quad \; + \;  \boldsymbol{\omega}^B\times\!\mathbf{J}\boldsymbol{\omega}^B,\\
\mathbf{f}^{B}_{\mathrm{cmd}}
&=
\mathbf{f}^{B}_{\mathrm{ff}}(w)
+\mathbf{R}\!\bigl(\mathbf q\bigr)^\top
\Bigl(
 -K_{l,1}\bigl(\mathbf p^I-\pi_p(w)\bigr)
- K_{l,2}\,\mathbf{v}^I
\Bigr),
\end{aligned}
\end{equation}
where the proportional and derivative gains $K_{r,1}$, $K_{r,2}$, $K_{l,1}$, and $K_{l,2}$ were selected empirically (see Table~\ref{tab:tsparams}). The complete control architecture is shown in Fig.~\ref{fig:ctrlschema}.
\begin{figure*}[htbp]
    \centering
    \includegraphics[width=0.8\linewidth]{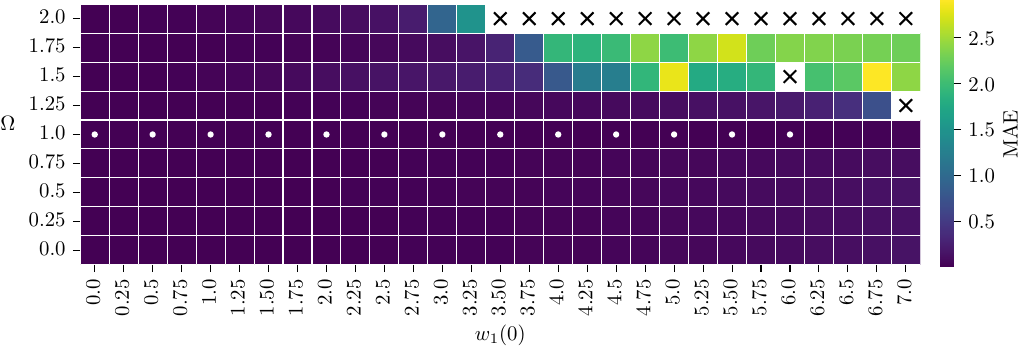}
    \caption{Grid of experiments over different exosystem configurations. The $x$-axis reports the initial condition $w_1(0)$, while the $y$-axis reports values of $\Omega$. Each cell is colored by the mean absolute vertical tracking error over a $30$\,s experiment. White dots highlight training configurations, while crosses indicate simulations that diverged due to instability and/or infeasibility.}
\label{fig:z-error-grid}
\end{figure*}
To assess the performance of our solution and its generalization capability to different exosystem models, we tested the neural controller under a wide range of simulation conditions, by varying both the exosystem frequency and the initial condition well beyond the configurations seen during training. We report the results in Fig.~\ref{fig:z-error-grid}. Each cell represents the mean absolute vertical tracking error over a $30$\,s experiment for a specific pair $(w_1(0),\Omega)$. Darker tones denote lower tracking error, while lighter tones correspond to larger errors. A cross inside a cell indicates that the simulation diverged at some point and we could not run the experiment for the full duration. Cells containing a white dot correspond to exosystem configurations that were used during the training phase of the PINN.
We can interpret the grid as a progression in difficulty: starting from the bottom-left corner, where both displacement and frequency are small, and moving towards the upper-right corner, where both are larger. 
\\
The PINN generalizes well across most of the domain, despite being trained on only a small subset of the tested configurations. This same training framework could be readily extended to a broader range of exosystem conditions. In the upper-right corner, where the most challenging scenarios are located, the tracking error increases and, in some cases, the simulation diverges. This behavior may be due in part to limitations of the current feedback design, but we want to emphasize that the design and tuning of an optimal stabilizing feedback law was not the primary objective of this work and was not further investigated.
\begin{figure}[htbp]
    \centering
    \includegraphics[width=1\linewidth]{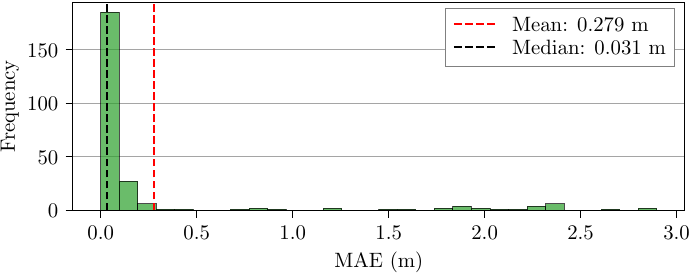}
    \caption{Histogram of the vertical tracking error over all grid experiments. The mean and median values are indicated by red and black dashed vertical lines, respectively.}
    \label{fig:estats}
\end{figure}
In Fig.~\ref{fig:estats} we report a statistical analysis of the vertical tracking error to further confirm that the error is negligible in most cases. The histogram shows the distribution of the mean absolute error over all grid experiments. The mean value is relatively high due to a small number of configurations with large error, but these appear as outliers; the median is on the order of $10^{-2}\,$m, which is representative of the typical behavior.
From the set of simulations, we selected two representative configurations to show in detail the temporal evolution of the error. One configuration corresponds to an in-distribution scenario, while the other is an out-of-distribution scenario, not shown at training time, with the goal of comparing the temporal behavior of the solution in both cases.
\\
The in-distribution case is shown in Fig.~\ref{fig:comparison-id}, while the out-of-distribution case is shown in Fig.~\ref{fig:comparison-ood}. In each figure, the top plot reports the trajectories of the exosystem reference signal and the vertical tracking error over time. The bottom plot zooms in on a selected time interval, highlighted in gray in the top plot, magnifying the error by a factor $100\times$ to make the residual ripple visible, which is otherwise not visible at the scale of the reference signal. In both scenarios, the error remains approximately two orders of magnitude smaller than the exogenous signal throughout the entire experiment. This further confirms that our controller is able to reconstruct the steady state manifold mappings and the feedforward control law that keep the motion close to the zero-error manifold.
It is worth noting that the residual oscillation does not exhibit a clear correlation with either the initial condition or the frequency of the exosystem, indicating that our PINN-based controller performs consistently over different configurations of the exosystem. This behavior supports the interpretation of the network as a learned operator that maps a family of dynamical systems into the corresponding solutions of the regulator equations, rather than solving a single fixed instance.
In Fig.~\ref{fig:3maps}, we show the solutions of the REs learned by the neural network, $c_b(w)$, $\pi_\phi(w)$, and $\pi_\theta(w)$, across different initial conditions of the exosystem with $w_1(0) \in [0,5]$\,m. We observe that the magnitude of the learned functions grows with the radius of the exosystem state, and that the mappings inherit the oscillatory behavior induced by the exosystem dynamics.
Furthermore, when the network receives as input the exosignal $w=(0,0)$, i.e., in the absence of external disturbance, it recovers the numerical value of the so-called \emph{trim condition} for the helicopter system, $(\pi_\phi = 0.044,\; \pi_\theta = 0.018,\; c_b = 0.0061)$~\cite{koo1998output}. These values correspond to the steady inputs required to keep the system hovering at a fixed altitude, providing an additional validation of the learned mappings.

\section{Discussion and Conclusions}

This work demonstrates that physics-informed neural networks can be effectively employed to solve the regulator equations for nonlinear output regulation. By enforcing the structure of the REs directly in the loss function, the neural network learns the steady state manifold and the corresponding feedforward input without requiring precomputed trajectories or datasets. The resulting model behaves as a solution operator, generalizing across a family of exosystem configurations rather than fitting a single instance as in classical approaches.
\\
The experimental results confirm that the learned operator reconstructs the zero-error manifold with high accuracy. The controller maintains small tracking error over a wide range of amplitudes and frequencies, including conditions not seen during training. 
The method is computationally efficient: the training stage requires only a few minutes on a consumer GPU, while the online phase consists of a lightweight forward pass, making real-time deployment feasible on embedded platforms.
\\
The main limitation of the approach is the full-information assumption, where the exosystem state $w$ is assumed to be available for feedback. Relaxing this requirement by moving to partial-information settings, or by assuming that $w$ is only accessible during training as privileged information and not at deployment, is a natural direction for future work.
A related avenue is to extend the framework from a single nominal plant to families of dynamical systems, for example by embedding parametric uncertainty in the regulator equations or randomizing plant parameters during training.  This would enable the learned operator to solve the output regulation problem for whole classes of plants. In parallel, integrating more advanced stabilizing feedback designs could further enlarge the feasible operating region and strengthen the stability guarantees.
\\
Overall, the results indicate that PINN-based controllers provide a viable and efficient tool for computing steady state solutions of nonlinear output regulation problems. The proposed methodology is general and can be extended to other nonlinear systems where solving PDEs is a central step in the control design.

\addtolength{\textheight}{-12cm}   



\appendix
The helicopter physical parameters used in the simulations are listed in Table~\ref{tab:phy-params}; see also~\cite{isidori2001computation}.
\begin{table}[h!]
    \centering
    \caption{Helicopter Physical Parameters}
    \begin{tabular}{r c l r c l r c l}
        $J_x$ & = & 0.14241 & $J_y$ & = & 0.27121 & $J_z$ & = & 0.2714 \\
        $\ell_M$ & = & -0.015 & $y_M$ & = & 0 & $h_M$ & = & 0.2943 \\
        $h_T$ & = & 0.1154 & $\ell_T$ & = & 0.8715 & $M$ & = & 4.9 \\
        $c_M^Q$ & = & 0.00445 & $D_M^Q$ & = & 0.6304 & $c_b^M$ & = & 25.23 \\
        $c_T^Q$ & = & 0.00506 & $D_T^Q$ & = & 0.00848 & $c_a^M$ & = & 25.23 \\
    \end{tabular}
    \label{tab:phy-params}
\end{table}
The main training hyperparameters and simulation settings are summarized in Table~\ref{tab:tsparams}.

\begin{table}[h!]
    \centering
    \caption{Training and Simulation Parameters}
    \label{tab:tsparams}

    \begin{tabular}{r c l l}
        \multicolumn{4}{c}{Training Parameters} \\[0.1cm]
        \hline \\[-0.2cm]
        $lr_{\text{init}}$  & = & $1\cdot10^{-3}$ & Initial learning rate \\
        $lr_{\text{final}}$ & = & $1\cdot10^{-6}$ & Final learning rate \\
        NN        & = & [32, 256, 256, 32] & Network architecture \\
        $n$                 & = & 24499         & Training samples \\
        $n_{\text{epochs}}$ & = & 100           & Training epochs \\
        $r$                 & $\in$ & $[0,6]$   & Sampling range of $w$ \\
    \end{tabular}

    \vspace{0.3cm}

    \begin{tabular}{r c l l}
        \multicolumn{4}{c}{Simulation Parameters} \\[0.1cm]
        \hline \\[-0.2cm]
        $\text{T}$ & = & $30\,\mathrm{s}$    & Simulation duration \\
        $dt$          & = & $0.01\,\mathrm{s}$  & Integration time step \\
        $K_{r,1}$      & = & $2$               & Rotational proportional gain \\
        $K_{r,2}$      & = & $0.1$             & Rotational derivative gain \\
        $K_{l,1}$      & = & $20$              & Translational proportional gain \\
        $K_{l,2}$      & = & $0.01$            & Translational derivative gain \\
    \end{tabular}
\end{table}

\bibliographystyle{IEEEtran}
\bibliography{bibliography}

\end{document}